\documentclass[a4paper,prd,english,aps,preprintnumbers,floats,onecolumn,nofootinbib]{revtex4}
\usepackage{graphicx}
\usepackage{subfig}
\usepackage[breaklinks,linktoc=page]{hyperref}
\usepackage{feynmp}
\usepackage{setspace}
\usepackage{soul}
\usepackage{enumerate}

\newcommand{\nc}{\newcommand}
\nc{\bea}{\begin{eqnarray}}
\nc{\eea}{\end{eqnarray}}
\nc{\be}[1]{\begin{equation} \mbox{$\label{#1}$}}
\nc{\ee}{\vspace{0.1cm}\end{equation}}
\nc{\eq}[1]{\mbox{Eq.\ (\ref{#1})}}
\nc{\fig}[1]{\mbox{Fig.\ (\ref{#1})}}
\nc{\ch}[1]{\mbox{Chapter\ \ref{#1}}}
\nc{\sect}[1]{\mbox{Section\ \ref{#1}}}
\nc{\HRule}{\rule{\linewidth}{0.5mm}}

\nc{\lh}{\lambda_h}
\nc{\ls}{\lambda_s}
\nc{\lhs}{\lambda_{hs}}
\nc{\lp}{\lambda_\phi}
\nc{\tdeps}{\tilde{\epsilon}}
\nc{\tdeta}{\tilde{\eta}}
\nc{\tdep}{\tilde{\epsilon}}
\nc{\tdu}{\tilde{U}}
\nc{\tdN}{\tilde{N}}
\nc{\tda}{\tilde{a}}
\nc{\tdH}{\tilde{H}}
\nc{\tdrho}{\tilde{\rho}}
\nc{\tdze}{\tilde{\zeta^2}}
\nc{\tdpartu}{\tilde{\partial_{\mu}}}
\nc{\tdpartd}{\tilde{\partial^{\mu}}}

\nc{\cv}{{\cal P}_{\zeta}(k)}

\nc{\bpo}{\bar{\phi_1}}
\nc{\bp}{\bar{\phi}}
\nc{\hdpo}{\hat{\delta\phi_1}}
\nc{\hdpt}{\hat{\delta\phi_2}}
\nc{\dpo}{\delta \phi_1}
\nc{\dpt}{\delta \phi_2}
\nc{\hdpi}{\hat{\delta\phi_i}}
\nc{\hdpj}{\hat{\delta\phi_j}}
\nc{\DoDo}{\partial_\mu (\hat{\delta \phi_1})\partial^\mu (\hat{\delta \phi_1})}
\nc{\DtDt}{\partial_\mu (\hat{\delta \phi_2})\partial^\mu (\hat{\delta \phi_2})}
\nc{\DoDt}{\partial_\mu (\hat{\delta \phi_1})\partial^\mu (\hat{\delta \phi_2})}
\nc{\xo}{\frac{\dpo}{\bpo}}
\nc{\xt}{\frac{\dpt}{\bpo}}

\nc{\hf}{\frac{1}{2}}
\nc{\hfrt}{\frac{1}{\sqrt{2}}}

\def\GeV{{\rm \ GeV}}

\begin{document}
\setlength{\unitlength}{1mm}
\title{Unitarity-violation in Generalized Higgs Inflation Models}
\author{Rose N. Lerner}
\email{rose.lerner@helsinki.fi}
\affiliation{Physics Department and Helsinki Institute of Physics, University of Helsinki, FIN-00014, Finland}
\author{John McDonald}
\email{j.mcdonald@lancaster.ac.uk}
\affiliation{Consortium for Fundamental Physics, Cosmology and Astroparticle Physics Group, Department of Physics, Lancaster University, LA1 4YB, UK}


\begin{abstract}
Unitarity-violation presents a challenge for non-minimally coupled models of inflation based on weak-scale particle physics. We examine the energy scale of tree-level unitarity-violation in scattering processes for generalized models with multiple scalar fields where the inflaton is either a singlet scalar or the Higgs. In the limit that the non-minimal couplings are all equal (e.g.\ in the case of Higgs or other complex inflaton), the scale of tree-level unitarity-violation matches the existing result. However if the inflaton is a singlet, and if it has a larger non-minimal coupling than other scalars in the model, then this hierarchy increases the scale of tree-level unitarity-violation. A sufficiently strong hierarchy pushes the scale of tree-level unitarity-violation above the Planck scale. We also discuss models which attempt to resolve the issue of unitarity-violation in Higgs Inflation.
\end{abstract}

\maketitle

\section{Introduction}

It is widely believed that cosmological inflation is due to a scalar field. Since the only fundamental scalar field in the Standard Model (SM) is the Higgs boson, the possibility that the Higgs boson could be the inflaton is very significant \cite{Bezrukov:2007ep}. Incorporating inflation into the SM opens up the possibility of directly relating inflation observables to experimental particle physics. Inflation with the Higgs boson is possible if the Higgs doublet has a large non-minimal coupling $\xi \sim 10^{4}$ to the Ricci scalar $R$ \cite{Bezrukov:2007ep}. Inflation based on a large non-minimal coupling was first proposed in \cite{spok} and later developed by \cite{salopek}. In this original scenario, the role of quantum corrections in connecting particle physics and inflation was first considered by \cite{Barvinsky:1994hx}. In the scenario of SM Higgs inflation, a lot of attention has been paid to calculating the predictions for inflation observables using quantum corrections \cite{Barvinsky:2008ia,barv1,barv2,newref2,newref3,Bezrukov:2009db}. Related models can also be constructed in well-motivated scalar extensions of the SM, in particular a gauge singlet scalar extension of the SM motivated by dark matter \cite{Clark:2009dc,Lerner:2009xg,Lerner:2011ge}. In this case, there are distinct predictions for the spectral index as a function of Higgs mass when inflation is along the singlet scalar direction, a model we refer to as $S$-inflation \cite{Lerner:2009xg,Lerner:2011ge}. (See also \cite{shafi}.)

Although these models are interesting, the large non-minimal coupling to gravity introduces a potentially dangerous side-effect. Graviton exchange in $2 \rightarrow 2$ scalar scattering causes unitarity-violation at tree-level. In today's vacuum, the energy scale of tree-level unitarity-violation is $\Lambda_0 \sim M_{p}/\xi$ \cite{Hertzberg:2010dc,burgess}. This is lower than the magnitude of the Higgs field at $N$ e-foldings before the end of inflation, $\phi_N \approx \sqrt{N/\xi}\,M_p$, and  approximately equal to the energy scale of the inflaton fluctuations (given by the Hubble parameter during inflation $H_*$). If this unitarity-violation is real, then the simplest assumption is that new particles and interactions should be introduced at or below the scale of unitarity-violation $\Lambda_0$ to restore unitarity and complete the theory. This new physics would then be expected to modify the Higgs potential at $\phi \gtrsim \Lambda_0$, making Higgs Inflation at least unpredictive and possibly ruling it out. The naturalness of the model is discussed in \cite{barv2}.

This simple interpretation of tree-level unitarity-violation is not the only possible interpretation. Other interpretations rely on the fact that the energy scale of tree-level unitarity-violation depends on the background value of the inflaton field $\bp$ \cite{Bezrukov:2010jz,Bezrukov:2011sz}.
For Higgs Inflation, this means that the scale of tree-level unitarity-violation is raised to $\Lambda_{inf} \sim M_p / \sqrt{\xi}$ during inflation \cite{Bezrukov:2010jz}. This scale is approximately equal to $\bp$ and larger than $H_\ast$. This has particular relevance to a recent proposal that the new physics which unitarizes Higgs Inflation could also depend on the background Higgs field during inflation \cite{Bezrukov:2011sz}. In this scenario, the scale of new physics during inflation, $M_{NP}$, could be as large as\footnote{It is important to distinguish between the scale of new physics $M_{NP}$, which is a mass scale associated with new particles introduced to complete the theory, and $\Lambda$, which is the energy at which unitarity is violated in scattering processes. In general, $M_{NP}$ must be less than  $\Lambda$.} the background-dependent scale of tree-level unitarity-violation \cite{Bezrukov:2011sz}. If $M_{NP}$ was substantially larger than the Higgs field during inflation, the new physics corrections to the Higgs potential may be small enough to leave inflation unaltered. However, this is not realised for the original model of Higgs Inflation because $\Lambda_{inf}(|\Phi|) \sim |\Phi|$, where $|\Phi|$ is the background Higgs field{\footnote{If the scale of new physics, $M_{NP}$, is exactly proportional to $|\Phi|$, the effective potential during inflation remains scale invariant and Higgs Inflation can occur \cite{Bezrukov:2011sz}. However, the link with weak scale observables is likely to be lost in this case.}}.

There have been a number of other suggestions to make the apparently unitarity-violating theory consistent while neither preventing inflation from occurring nor rendering the theory unpredictive. In \cite{Lerner:2010mq} it was proposed that the non-minimal coupling is accompanied by additional non-renormalizable Higgs interactions which cancel unitarity-violation. In this case no new fields are required. However, it remains to be shown whether such an approach can be made consistent with quantum corrections and the effect of additional potential and Yukawa interactions \cite{Lerner:2010mq}. In \cite{Giudice:2010ka} it was claimed that unitarity-violation in Higgs Inflation could be cancelled by the addition of a gauge singlet scalar to the SM with a particular form of scalar potential. This would provide an example of a new physics sector which completes the theory. Moreover, it would provide an example of a new physics sector where the new particle masses are explicitly background-dependent. However, we will show that this model is essentially the addition to the SM of an induced gravity inflation model with a mass scale much larger than the weak scale  and therefore not a true completion of Higgs Inflation.

Another suggestion is that strong coupling in graviton-exchange processes may unitarize the cross-section without requiring new physics. If this turns out to be possible (which is not known at present), it would provide a simple mechanism for unitarity-conservation which is naturally background-dependent.  Strong coupling could, in principle, result in a unitary high-energy scattering cross-section, even though unitarity is violated in the tree-level process \cite{Bezrukov:2009db}. Tree-level breakdown of unitarity is not proof of actual unitarity-violation and, in general, perturbation theory will break down before the energy of tree-level unitarity-violation is reached. This possibility was first noted in \cite{willenbrock} and has been discussed more recently in \cite{dono}. Because strong coupling unitarization introduces no new particles or interactions, the model would remain unaltered. As the scale of tree-level unitarity-violation is now interpreted as a physical strong coupling scale, the scale automatically takes a value determined by the Higgs background during inflation. Therefore Higgs Inflation would be unaltered (up to possible modifications of the radiative corrections \cite{Bezrukov:2009db})  provided that the relevant energy scale during inflation, the Hubble parameter $H_\ast$, is less than the background-dependent strong coupling scale $\Lambda$.

The phenomenological validity of Higgs Inflation models, by which we mean that inflation can be consistently analysed independently of the solution to the unitarity problem, depends on the energy scale of tree-level unitarity-violation. This scale is known both for pure Higgs Inflation and for gauge-singlet scalar models with a single non-minimal coupling. However, it is not known in the generalised case with multiple scalars and with different non-minimal couplings. The main objective of this paper is to estimate the scale of tree-level unitarity-violation in a range of non-minimally coupled models of inflation. We will study a generalized model with multiple scalar fields, each with its own non-minimal coupling to gravity.  This generalized model will contain the scalar sector of both Higgs Inflation and S-inflation as limiting cases. In this paper we focus on tree-level unitarity-violation. We emphasize the absence of tree-level unitarity-violation can only provide a sufficient condition for unitarity-conservation at energy scales where perturbation theory is clearly valid.  A necessary condition for unitarity-conservation would require a complete non-perturbative understanding of unitarity-violation in scattering processes.

We will focus on tree-level unitarity-violation coming directly from the non-renormalizable coupling. This results from two-particle scattering via graviton exchange in the Jordan frame. In the Einstein frame this is equivalent to scattering via a non-renormalizable interaction in a minimally coupled theory with the potential set to zero.  The effect of including the potential in the Jordan frame may be considered secondary, corresponding to graviton exchange between the non-minimal coupling and the potential interactions. As we will discuss, only graviton-exchange occurring directly via the non-minimal coupling leads unambiguously to dangerous tree-level unitarity-violation in scattering processes.

One of our main results is that the scale of tree-level unitarity-violation depends on all of the non-minimal couplings, not just the non-minimal coupling of the inflaton. In particular, if the inflaton is a real scalar, the scale of tree-level unitarity-violation can be much larger than in the case of Higgs Inflation if there is a hierarchy of non-minimal couplings. (No such hierarchy is possible in Higgs Inflation since the Goldstone bosons and Higgs boson in the Higgs doublet all have the same non-minimal coupling.) This allows the construction of inflation models where the scale of tree-level unitarity-violation is generally as large as the effective Planck scale, even in the present vacuum with no large background field.

    Our paper is organised as follows. In \sect{HS} we review non-minimally coupled inflation models based on the SM and on its extensions to include a gauge singlet scalar.  In \sect{UV} we present the generalized model and its Einstein frame expansion, both in the presence of a background inflaton field and in the vacuum. Scattering cross-sections for $2 \rightarrow 2$ processes via the non-minimal coupling are computed and the energy scale of tree-level unitarity-violation is obtained as a function of the non-minimal couplings. In \sect{pot} we consider the effect of the non-polynomial Einstein frame potential in the presence of a background field. In \sect{dis} we discuss the possibility of unitarity-conservation both by a background-dependent scale of new physics and by strong coupling. We also discuss the Giudice-Lee model \cite{Giudice:2010ka} and apply our general results to derive the scale of tree-level unitarity-violation in this model and to clarify its true nature. In \sect{conc} we present our conclusions.

\section{Inflation From Weak-Scale Particle Physics: Higgs Inflation and $S$-Inflation}

\label{HS}
The Lagrangian of the original Higgs Inflation model is the SM Lagrangian plus a non-minimal coupling of $H^\dagger H$ to the Ricci scalar $R$. This coupling is generally expected to exist,  being generated by quantum corrections in curved space. The Jordan frame action is given by \cite{Bezrukov:2007ep}
\be{JHiggs}
 S_H  =  \int \sqrt{-\!g}\,d^4\! x \left[{\cal L}_{\overline{SM}} + \left(D_\mu H\right)^{\dagger}\left(D^\mu H\right) - \frac{M^2R}{2} - \xi_h H^{\dagger}H R -  \lh \left(H^{\dagger}H - \frac{v^2}{2}\right)^2\right]
\ee
where ${\cal L}_{\overline{SM}}$ is the Standard Model Lagrangian density minus the purely Higgs doublet terms.

An alternative model, $S$-inflation \cite{Lerner:2009xg,Lerner:2011ge}, is based on the simplest scalar extension of the SM, namely the addition of a gauge singlet scalar, which may be real or complex. For appropriate values of the coupling of the singlet to the Higgs, this is the simplest extension of the SM which is able to account for WIMP dark matter \cite{gsdm1}. In $S$-inflation the non-minimally coupled singlet is also the inflaton and is assumed to have a large non-minimal coupling $\xi_s$. This field is coupled to the Higgs doublet, which is assumed to have a smaller non-minimal coupling $\xi_h \ll \xi_s$. The Jordan frame action is given by \cite{Lerner:2009xg,Lerner:2011ge}
\bea
\label{SHiggs}
 S_S & = & \int \sqrt{-\!g}\,d^4\! x \Bigg[{\cal L}_{\overline{SM}} + \left(\partial_\mu S\right)^{\dagger}\left(\partial^\mu S\right) + \left(D_\mu H\right)^{\dagger}\left(D^\mu H\right) - \frac{M^2R}{2} - \xi_s S^{\dagger}SR - \xi_h H^{\dagger}H R \nonumber \\
 & &  - m_s^2S^{\dagger}S -  \ls \left(S^{\dagger}S\right)^2  - \lhs S^{\dagger}S H^{\dagger}H - \lh \left(H^{\dagger}H - \frac{v^2}{2}\right)^2\Bigg].
\eea
If $\xi_{s} \gg \xi_{h}$ then inflation will be along the $S$ direction, which minimises the potential in the Einstein frame. The coupling $\lhs$ can be determined by the requirement that $S$ produces the observed density of dark matter, whereas $\ls$  is a free parameter, bounded by vacuum stability and perturbativity \cite{Lerner:2009xg,Lerner:2011ge}.

\subsection{Formalism for inflation}

We will calculate scattering amplitudes in the Einstein frame, which is related to the Jordan frame (where the theories are physically defined) by a conformal transformation. Quantities in the Einstein frame are marked with a tilde. Throughout this paper, $\phi_1$ will denote the inflaton. The Einstein frame metric is given by
\be{gJE}
\tilde{g}_{\mu\nu} = \Omega^{2}(\phi_i) g_{\mu\nu}
~,\ee
where $\Omega ^2(\phi_i)$ is a conformal factor that depends on all of the scalar fields. During inflation, $\Omega^2 \simeq \xi_1 \phi_1^2/M_p^2 \gg 1$. Energy scales in the Einstein frame are transformed to the Jordan frame via
\be{rel}
\tilde{E} = \frac{E}{\Omega} ~.
\ee
At $N$ e-foldings before the end of inflation, the inflaton field has a value
\be{phiN}
\phi_{N} \simeq \sqrt{\frac{N}{\xi_1}}M_p ~.
\ee
Note that $\phi_N$ is generally less than the effective Planck scale in the Jordan frame during inflation, given by
\be{effMp}
M_{eff}^2 = M_p^2 + \xi_1 \phi_N^2 \approx \xi_1 \phi_N^2   ~.
\ee
The Hubble scale during inflation, $H_{\ast}$, in the Jordan frame is determined by the magnitude of the observed density perturbation,
\be{Hstar}
{H}_{\ast} \simeq \sqrt{\frac{\lambda}{3}} \frac{M_p}{2\xi_{1}} \approx 2\times 10^{-5} M_p
~,\ee
where $\lambda$ is the quartic self coupling of the inflaton. The ratio $\sqrt{\lambda}/\xi_1$ is therefore fixed by the amplitude of the perturbations. The coupling $\lambda$ is fixed by SM phenomenology to be ${\cal O}(0.1)$ for Higgs Inflation but could be substantially lower in more general models, such as $S$-inflation, in which case smaller $\xi_1$ is possible. The Hubble scale gives the energy of the quantum fluctuations of the scalar fields about the background field. Therefore, a necessary (but not sufficient) condition for a viable model is that $H_*$ must be less than the energy scale of unitarity-violation during inflation.

We note that the equivalence of the theory in the Jordan and Einstein frame is a subtle issue (see e.g.\ \cite{barv1,barv2,flan}). The conformal transformation from the Jordan to Einstein frame is not simply a rescaling but also a mixing of the gravitational and scalar degrees of freedom, as is clear from \eq{gJE}. Therefore the equivalence of the quantized theory in the different reference frames will depend upon which degrees of freedom are quantized, in particular whether gravitational degrees of freedom are quantized \cite{barv1,barv2,flan}. However, because we are considering tree-level scattering, the equivalence theorem of the scattering matrix under non-linear local field redefinitions ensures that the cross-sections can be calculated in either frame (see Section IIc of Ref. \cite{flan} and references therein).

\section{Generalized Higgs Inflation Model and tree-level unitarity-violation }
\label{UV}

In this section we will study a generalized Higgs Inflation model with multiple real scalar fields, each with its own non-minimal coupling to gravity. Our goal is to estimate the scale of tree-level unitarity-violation  in 2 $\rightarrow$ 2 elastic scattering processes due to non-potential terms in the Einstein frame. We do this both in the vacuum and in the presence of a large background field. We do not explicitly include gauge fields, but we do discuss how they modify the results from the scalar field case. Our method is to use dimensional analysis to identify the leading tree-level  unitarity-violating processes and then to confirm the estimate of $\tilde{\Lambda}$ by a full calculation of the cross-section.

The generalized Higgs Inflation model is defined in the Jordan frame by
\be{j0}
 S_J = \int \sqrt{-\!g}\,d^4\! x \Big(\hf\partial_\mu \phi_i \partial^\mu \phi_i - V\left(\phi_{i}\right) - \hf M_p^2R - \hf \xi_i \phi_i^2 R  \Big)    ~,
\ee
where a sum over $i$ is assumed throughout this section. In the Einstein frame this becomes
\bea \label{j1}
S_E &=& \int d^4x\sqrt{-\tilde{g}}  \left[\hf \left(\frac{\Omega^2 + \frac{6\xi_i^2 \phi_i^2}{M_p^2}}{\Omega^4}\right)\tilde{g}^{\mu\nu}\partial_\mu \phi_i \partial_\nu \phi_i + \sum_{j<i}\frac{6\xi_i\xi_j\;\phi_i\;\phi_j \; \tilde{g}^{\mu\nu}\;\partial_\mu \phi_i\partial_\nu \phi_j}{M_p^2\Omega^4}  -
\frac{V(\phi_{i})}{\Omega^4} -\hf M_P^2\tilde{R} \right] ~,
\eea
where the conformal factor is
\be{j2}
\Omega^2 = 1 + \frac{\xi_i \phi_i^2}{M_p^2}   ~.\ee

 The second term in \eq{j1} is due purely to the non-minimal coupling in the Jordan frame, and is the primary source of tree-level unitarity-violation in Higgs Inflation-type models. This is a non-renormalizable term with positive powers of the scalar fields in the limit where $\Omega \rightarrow 1$. Therefore it unambiguously violates tree-level unitarity. (The energy scale of tree-level unitarity-violation in the vacuum for pure Higgs Inflation is $M_{P}/\xi$, which is well below the field strength at which $\Omega$ deviates from 1.)
In contrast, the potential term in the Einstein frame is proportional to $1/\Omega^4$. This introduces {\it inverse} powers of the scalar fields in the limit of large field strengths. This is the opposite behaviour from the non-renormalizable terms which usually violate unitarity, which have positive powers of the fields. Therefore it is unclear whether such terms can violate unitarity in scattering processes. (Scattering due a non-polynomial potential, such as $V/\Omega^4$, is an unsolved problem in field theory.) Similar comments apply to the possibility of unitarity-violation via Yukawa couplings to fermions. For this reason we will focus primarily on unitarity-violation due purely to the non-minimal coupling. We will, however, consider unitarity-violation due to the non-polynomial potential when expanded about a large background field, which can be understood using conventional methods.

\subsection{One Real Scalar}

\label{wrr}

We first consider the case of only one real scalar and no potential. In this case the Einstein frame action with $V = 0$ is
\be{rsE}
S_E  =  \int d^4x\sqrt{-\tilde{g}} \left(- \frac{M_P^2\tilde{R}}{2} + \frac{1}{2} \left(\frac{M_p^2 + \xi_1 \phi_1^2 + 6\xi_1^2 \phi_1^2}{(M_p^2 + \xi_1 \phi_1^2)^2}\right)\partial_\mu \phi_1 \partial^\mu \phi_1 \right)  ~.\ee
As the kinetic term contains only one dynamical field, it can be fully canonically normalised by a field redefinition $\phi_{1} \rightarrow \chi$, where
\be{j3}
\frac{d\chi}{d \phi_{1}} = \sqrt{\frac{\Omega ^2 + 6 \xi ^2\phi_{1}^{2}/M_P^2}{\Omega ^4}}     ~,
\ee
in which case the action becomes
\be{rsE2} S_E  = \int d^4x\sqrt{-\tilde{g}} \left(- \frac{M_P^2\tilde{R}}{2} + \hf\partial_\mu \chi \partial^\mu \chi \right)
~.\ee
Therefore, there is no tree-level unitarity-violation in 2 $\rightarrow 2$ scattering processes in a non-minimally coupled theory with $V = 0$ and only one real scalar (as first noted in \cite{Lerner:2009na}). In the Jordan frame, dimensional analysis appears to give unitarity-violation \cite{Burgess:2009ea,Barbon:2009ya}. However this analysis misses the cancellation of $s$, $t$ and $u$-diagrams \cite{toms,Hertzberg:2010dc}. The Jordan frame cancellation is equivalent to a similar cancellation between momentum factors in the Einstein frame when calculating with $\phi_1$ rather than with $\chi$. This is true both in the vacuum and at large background fields.

\subsection{Two Real Scalars}

  We do not expect the same cancellation in the case with multiple scalars because it is not possible to obtain a canonically normalised theory in the Einstein frame. This corresponds to the absence of s-channel graviton exchange diagrams in the Jordan frame when there are two different scalars in the initial state, preventing the s-, t-, u-diagram cancellation.

The Einstein frame action for two real scalars when $V = 0$ is
\bea \label{Eaction}
S_E = \int d^4x\sqrt{-\tilde{g}}  \left[{\cal L}_{11} + {\cal L}_{22} + {\cal L}_{12}  -\hf M_P^2\tilde{R} \right],
\eea
where
\be{Lii}
 {\cal L}_{ii} = \hf \left(\frac{\Omega^2+\frac{6\xi_i^2\phi_i^2}{M_p^2}}{\Omega^4}\right) \tilde{g}^{\mu\nu}\partial_\mu \phi_i \partial_\nu \phi_i
\ee
and
\be{L12}
{\cal L}_{12} = \frac{6\xi_1\xi_2\;\phi_1\;\phi_2 \; \tilde{g}^{\mu\nu}\;\partial_\mu \phi_1\partial_\nu \phi_2}{M_p^2\Omega^4}
~.\ee
During inflation, $\phi_1$ has a large background value ($\bpo \approx \sqrt{N}M_p/\sqrt{\xi_1}$) whereas $\phi_2$ has zero background value. The fields can then be written as
\be{exp}
\phi_1 = \bpo + \dpo,\;\;\;\;\phi_2 = \dpt
\ee
where $\dpo,\dpt \ll \bpo$.

\subsubsection{Scale of tree-level unitarity-violation in vacuum}
\label{vac}
 The scale of tree-level unitarity-violation in today's vacuum ($\bpo \simeq 0$) can be read directly{\footnote{For $\bpo \ll M_{Pl}/\sqrt{\xi_{1}}$  the conformal factor $\Omega \approx 1$ and the field is canonically normalised in either the Jordan or Einstein frame.}} from the interaction terms in \eq{Eaction} by taking the limit $\Omega \rightarrow 1$. This gives
\be{Lam_0_12}
\Lambda_{0} \approx \frac{M_p}{\sqrt{\xi_1 \xi_2}} ~,
\ee
which reduces to the well-known result $\Lambda_0 \approx M_p/\xi_1$ for $\xi_1 = \xi_2$.  On the other hand, if $\xi_2 \lesssim 1/\xi_1$ then $\Lambda_0$ would become larger than $M_p$ and it would be possible to have $S$-inflation without any dangerous tree-level unitarity-violation below the scale of quantum gravity. The physical explanation for this is that in the case of small $\xi_2$, the theory approaches the pure singlet model, which has canonical kinetic terms in the Einstein frame and so no unitarity-violation. However, because $\xi_1$ and $\xi_2$ are coupled via their renormalization group equations, this hierarchy will only be natural with respect to quantum corrections if it is protected by a symmetry. In particular, supersymmetry can protect non-minimal couplings from quantum corrections \cite{tj}. However, supersymmetry would imply a complex singlet scalar, whereas we require a real singlet scalar here.

\subsubsection{Scale of tree-level unitarity-violation in a large background field}

To obtain the general scale of tree-level unitarity-violation in a large background field, we expand ${\cal L}_{11}$, ${\cal L}_{22}$ and ${\cal L}_{12}$
in terms of $\dpo$ and $\dpt$. The resulting action for $\dpo$ and $\dpt$ will not be canonically normalized. We therefore define canonically normalized fields $\varphi_1$ and $\varphi_2$. In the limit $\bpo \gg M_{p}/\sqrt{\xi_1}$, these are given by
\be{can1}
\varphi_1 = \frac{\sqrt{6}M_p}{\bpo} \dpo
\ee
and
\be{can2}
\varphi_2 = \frac{M_p}{\sqrt{\xi_1}\bpo} \dpt ~. \ee
Keeping only terms which can in principle contribute to two-particle elastic scattering, we find
\be{L11_b}
{\cal L}_{11} \simeq \hf \partial_\mu \varphi_1 \partial^\mu \varphi_1 - \frac{\xi_2}{M_p^2} \varphi_2^2 \partial_\mu \varphi_1 \partial^\mu \varphi_1,
\ee
\be{L22_b}
{\cal L}_{22} \simeq \hf \partial_\mu \varphi_2 \partial^\mu \varphi_2 +\left(\frac{\varphi_1^2}{4M_p^2}- \frac{\varphi_1}{\sqrt{6}M_p}  \right) \partial_\mu \varphi_2 \partial^\mu \varphi_2
\ee
and
\be{L12_b}
{\cal L}_{12} \simeq \frac{\sqrt{6}\xi_2}{M_p}\varphi_2 \partial_\mu \varphi_1 \partial^\mu \varphi_2 - \frac{3\xi_2}{M_p^2}\varphi_1 \varphi_2 \partial_\mu \varphi_1 \partial^\mu \varphi_2.
\ee
These expressions are only valid during inflation, when $\bpo \gg M_p / \sqrt{\xi_1}$.

   Using the interactions \eq{L11_b}, \eq{L22_b} and \eq{L12_b}, we can obtain the scale of tree-level unitarity-violation via dimensional analysis ($\tilde{\Lambda}_{dim}$), by introducing appropriate factors of $\tilde{E}$ to make the coefficient of the interaction terms in ${\cal L}$ dimensionless. The unitarity-violating scale is then given by the magnitude of $\tilde{E}$ for which the coefficient is of order unity, since a unitarity-conserving $2\rightarrow 2$ scattering process must have a dimensionless amplitude less than ${\cal O}(1)$. Therefore this procedure gives a good estimate of the energy scale of unitarity-violation due to each Einstein-frame interaction term, in the absence of possible cancellations between Feynman diagrams.

    Before proceeding, we comment on the interpretation of the unitarity-violation scale and the corresponding scale of new physics. We can define the unitarity-violation scale in either the Einstein or the Jordan frame. However, physics is defined in the Jordan frame. If unitarity is restored by a sector of new particles with mass $M_{NP}$ then we require that $M_{NP}$ is less than the unitarity-violation scale in the Jordan frame,  $M_{NP} < \Lambda(\bpo)$. The new physics that restores unitarity is expected to introduce interactions of the form $\phi^{4+2n}/M_{NP}^{2n}$. Therefore a minimal requirement for a consistent inflation model is that $\bpo < M_{NP} \lesssim \Lambda(\bpo)$ in the Jordan frame during inflation, where $M_{NP}$ may depend\footnote{If scale-invariance is exact at large $\bpo$ then $M_{NP} \propto \bpo$. In this case inflation can be achieved for any $M_{NP}$, since scale-invariance implies that $V \propto \bpo^4$ for the full potential, although any link with weak-scale physics is lost \cite{Bezrukov:2011sz}.} on $\bpo$.

\begin{table}
\begin{tabular}{|l|c|c|c|c|c|}
  \hline
   & A& B&C &D  &E   \\ \hline
   Lagrangian term & $-\frac{\xi_2 \varphi_2^2}{M_p^2} \partial_\mu \varphi_1 \partial^\mu \varphi_1$& $-\frac{\varphi_1}{\sqrt{6}M_p} \partial_\mu \varphi_2 \partial^\mu \varphi_2$ & $\frac{\varphi_1^2}{4M_p^2}  \partial_\mu \varphi_2 \partial^\mu \varphi_2$ & $ \frac{\sqrt{6}\xi_2}{M_p}\varphi_2 \partial_\mu \varphi_1 \partial^\mu \varphi_2$&$-\frac{3\xi_2}{M_p^2}\varphi_1 \varphi_2 \partial_\mu \varphi_1 \partial^\mu \varphi_2$  \\ \hline
Einstein: $\tilde{\Lambda}_{dim}$& $M_p/\sqrt{\xi_2}$ & $M_p$ & $M_p$ & $M_p/\xi_2 $ & $M_p/\sqrt{\xi_2}$\\
\hline
Jordan: $\Lambda_{dim}$ & $\sqrt{\xi_1/\xi_2}\;\bpo$ & $\sqrt{\xi_1} \;\bpo$  &  $\sqrt{\xi_1} \;\bpo$  &  $\sqrt{\xi_1}/\xi_{2} \;\bpo$  &
$\sqrt{\xi_1/\xi_2}\;\bpo$  \\
\hline
  $\Lambda_{dim} (\xi_1 \gg \xi_2 \lesssim 1)$ & $\Lambda_{dim} \gtrsim   M_{eff}$ & $\Lambda_{dim} \approx M_{eff}$ & $\Lambda_{dim} \approx M_{eff}$ & $\Lambda_{dim} \gtrsim M_{eff}$ & $\Lambda_{dim} \gtrsim M_{eff}$ \\
\hline $\Lambda_{dim} (\xi_2 \simeq \xi_1)$ & $\Lambda_{dim} \ll M_{eff}$ & $\Lambda_{dim} \approx M_{eff}$ &  $\Lambda_{dim} \approx M_{eff}$ & $\Lambda_{dim} \ll M_{eff}$ &$ \Lambda_{dim} \ll M_{eff}$ \\ \hline
\end{tabular}
\caption{Scale of tree-level unitarity-violation from dimensional analysis. The final rows of the table show the Jordan frame tree-level unitarity-violation scale in the limits $\xi_1 \gg \xi_2 \approx 1$ and $\xi_2 \simeq \xi_1$. $M_{eff} \approx \sqrt{\xi_1} \bpo$ is the effective Planck scale in the Jordan frame during inflation. } \label{table1}
\end{table}

In Table~\ref{table1} we show the Einstein frame background-dependent tree-level unitarity-violation scale $\tilde{\Lambda}_{dim}$, where the corresponding interaction terms are labelled for later convenience. We also show the corresponding Jordan frame scale ($\Lambda_{dim} = \Omega \tilde{\Lambda}_{dim}$) in the limits $\xi_1 \gg \xi_2$ and $\xi_1 \simeq \xi_2$, relevant to $S$-inflation and Higgs Inflation respectively. Unless there are cancellations between diagrams, interaction D will give the lowest tree-level unitarity-violation scale. We next compute the exact energy of tree-level unitarity-violation in elastic scattering for this process.

There are two elastic scattering processes to consider. The first, $\varphi_2 \varphi_2 \rightarrow \varphi_2 \varphi_2 $, can occur via $\phi_1$ exchange. In this case we find a $s$-, $t$-, $u$-cancellation, just as in the singlet case.
In contrast, there is no cancellation for the process $\varphi_1 \varphi_2 \rightarrow \varphi_1 \varphi_2$. The matrix element from $\varphi_1$ exchange via interaction D is
\be{M_D}
{\cal M}_D = -\frac{6i\xi_2^2 \tilde{E}^2}{M_p^2}
~.\ee
The optical theorem condition for unitarity-conservation in elastic scattering is \cite{itz,willenbrock}
\be{uc1}
\left|{\rm Re}(a_l)\right| \leq \hf
~,\ee
where the partial wave amplitudes $a_l$ are given by
\be{uc2}
-i{\cal M} = 16\pi\sum_{l=0}^{\infty}(2l+1)P_l(\cos\theta) a_l
~.\ee
Thus at tree-level,
\be{a0}
a_0 = \frac{3\xi_2^2\tilde{E}^2}{8\pi M_p^2}
\ee
and the tree-level unitarity-violation scale in the Einstein frame is
\be{uc3}
\tilde{E} \leq \tilde{\Lambda}_{12} = \sqrt{\frac{4\pi}{3}}\frac{M_p}{\xi_2}
~. \ee
This corresponds to the Jordan frame scale
\be{uc3J}
\Lambda_{12} = \sqrt{\frac{4\pi\xi_1}{3\xi_2^2}} \bpo
~. \ee
This imposes the strongest constraint in the case of two real scalars and agrees with the dimensional estimate.

In general, $\Lambda_{12}$ will become larger than $\bpo$ when $\xi_2 \lesssim \sqrt{\xi_{1}}$, in which case a $\bpo$-dependent scale of new physics could allow inflation to be unaffected by corrections due to unitarity-conserving new physics. Alternatively, if $\Lambda_{12}$ represents a background-dependent strong coupling scale, then $\Lambda_{12}$ is larger than $H_{*} (\approx M_{Pl}/\xi_{1})$ if $\xi_{2} \lesssim \xi_{1}^{3/2}$.

We next apply this general result (\eq{uc3J}) to four specific cases.
\begin{enumerate}[(i)]
\item {\bf S-inflation with a real inflaton}: If $\xi_1 \gg \xi_2 \lesssim 1$, the scale of tree-level unitarity-violation is approximately equal to the effective Planck mass, $M_{eff} \approx \sqrt{\xi_1} \bpo$, throughout inflation. This scenario is possible in $S$-inflation and other models with a singlet inflaton. As a result, tree-level unitarity-violation is effectively eliminated from the low-energy effective theory during inflation and appears only at the scale where gravity itself is expected to violate unitarity.
\item {\bf $S$-inflation with a complex inflaton}: This is described by $\xi_2 = \xi_1$. The Jordan frame scale in this case is $\Lambda_{dim} \approx \bpo / \sqrt{\xi_1}$. This is much lower than the background field $\bpo$ and similar to the Hubble scale $H_\ast$ in the Jordan frame. In the case where the two scalars are physical particles, this would imply large corrections to inflationary perturbations from the new physics of unitarity-conservation. In the case of a strong coupling interpretation of $\Lambda_{dim}$, large corrections to inflation observables from strong coupling may occur because $H_* \sim \Lambda_{dim}$ \cite{Bezrukov:2009db}. In the case of new physics, the corrections at $\bpo \gg \Lambda_{dim}$ would be expected to strongly modify the inflaton potential and possibly rule out the model. Therefore $S$-inflation with a complex singlet scalar is disfavoured if $\Lambda$ represents a true breakdown of unitarity-violation, and is likely to make the model unpredictive in the case where $\Lambda$ is a strong coupling scale.
\item {\bf Higgs Inflation:}  The case $\xi_1 = \xi_2$ is relevant to Higgs Inflation at scattering energies for which the Goldstone bosons may be considered physical. However, the Goldstone scalars in the Higgs doublet can only be considered physical at scattering energies larger than the gauge boson masses. In this case the effective unitarity-violation scale in the Jordan frame will be approximately given by the gauge bosons masses, $\Lambda \approx g\bpo$. This is because at smaller scattering energies the massive gauge bosons will decouple and the effective theory will reduce to a single real Higgs scalar, which conserves unitarity\footnote{This conclusion agrees with the results of \cite{Bezrukov:2010jz}, where it is argued that the decoupling of the Higgs mode due to the non-minimal coupling implies that unitarity must be violated at energies greater than the gauge boson masses.}. However, there could still be dangerous corrections to the inflaton potential, because $\Lambda \approx g \bpo \lesssim \bpo$.
\item {\bf Higgs Inflation with an additional scalar:} Additional scalar fields could decrease the scale of tree-level unitarity-violation in models where the Higgs boson is the inflaton, such as the model in \cite{Clark:2009dc}, where an additional singlet is added to Higgs Inflation. This is because in pure Higgs Inflation, tree-level unitarity-violation in a large background Higgs field occurs only above the mass of the gauge bosons, but with an additional non-minimally coupled gauge singlet scalar, tree-level unitarity-violation can occur at a lower energy via scattering of the physical Higgs scalar from the singlet scalar.
This effect is strongest when the non-minimal couplings are equal.
\end{enumerate}

\section{Effect of the Potential}
\label{pot}
In \sect{UV} we considered tree-level unitarity-violation in the limit $V = 0$. In this case, tree-level unitarity-violation is due to graviton exchange via the non-minimal coupling. We next consider the effect of the potential. When $V \neq 0$, it is possible to have graviton exchange scattering processes between the non-minimal coupling $\phi^2 R$ and the potential term $\sqrt{-g} V$. This will produce $\lambda$-dependent contributions to scattering cross-sections. We take the same approach as in the previous sections and expand the potential around a large background value in terms of the canonically normalised fields. Because many-particle processes are involved, we obtain unitarity-violation scales by dimensional analysis only. As in the previous sections, we begin with the case of a single real scalar. We first explain our method for estimating the scale of unitarity-violation. This method is only applicable when expanding around a large background value --- in the vacuum, it is not yet understood how to analyse the non-polynomial potential in a perturbative form\footnote{Previous discussions (e.g. \cite{Bezrukov:2010jz,Barbon:2009ya}) have considered the expansion of the Einstein frame potential around the scale $M_{P}/\sqrt{\xi}$ and unitarity-violation from individual terms in the expansion. However, such an expansion is unlikely to be valid when the energy is comparable with the expansion scale. The complete potential should be considered in this case.}.

To calculate non-renormalizable potential interactions in the Einstein frame, we need to consider $2 \rightarrow n$ inelastic scattering processes. In this case we can use the optical theorem to estimate the unitarity-violation scale. The optical theorem gives the total cross section to be \cite{itz}
\be{op1}   \sigma_{TOT}  = \frac{ {\rm Im} \left[ A(\theta = 0) \right]}{s}    ~,\ee
where $A(\theta) = -i {\cal M}(\theta)$ is the $2 \rightarrow 2$ elastic scattering amplitude, given by \eq{uc2}. Thus if we assume that the total elastic cross-section is dominated by the $a_{0}$ term, as in the case of $2 \rightarrow 2$ scalar scattering (or, more generally, if we assume the elastic cross-section is dominated by the low multipoles) then
\be{op2}  \sigma_{TOT} \approx \frac{ 16 \pi {\rm Im}(a_{0})}{s}   \leq \frac{ 4 \pi}{\tilde{E}^2} ~,\ee
where we have used the unitarity bound $0 \leq {\rm Im}(a_{0}) \leq 1$ and $s = 4\tilde{E}^2$.
The total cross-section is estimated dimensionally from the non-renormalizable coupling. If the coupling has the form $\phi^n/M^{n-4}$, where
$M$ has dimensions of mass, then dimensionally
\be{op3} \sigma_{TOT} \sim  \frac{\tilde{E}^{2\left(n-5\right)}}{M^{2\left(n-4\right)}}   ~.\ee
Therefore the dimensional unitarity bound on $\tilde{E}$ is
\be{op4} \tilde{E} \lesssim  M   ~.\ee
The actual bound could be somewhat larger once phase space factors are included, but we expect by only an order of magnitude at most.

\subsection{Potential sector: single real scalar}
In \sect{wrr}, we showed that there is no unitarity-violation in the case of one real scalar when $V=0$. We now consider the effect of the potential $V(\phi_1)$. In the Einstein frame this becomes a non-polynomial potential given by
\be{nonpoly}
\tilde{V}(\phi_1) \equiv \frac{V\left(\phi_{1}\right)}{\Omega^{4}}
= \frac{\frac{1}{4}\lambda\phi_1^4}{\left(1+\xi_1 \phi_1^2/M_p^2\right)^2}.
\ee
Expanding around a large background field value $\bpo$ we find
\be{Vsinglet}
\tilde{V}(\dpo) \approx \frac{\lambda M_p^4}{4\xi_1^2} \left[ 1 + \frac{M_p^2}{\xi_1 \bpo^2}\left(4\left(\frac{\dpo}{\bpo}\right) - 6\left(\frac{\dpo}{\bpo}\right)^2 + 8\left(\frac{\dpo}{\bpo}\right)^3 - 10\left(\frac{\dpo}{\bpo}\right)^4 + 12 \left(\frac{\dpo}{\bpo}\right)^5 - 14\left(\frac{\dpo}{\bpo}\right)^6 +\cdots\right)   \right].
\ee
Terms which are higher order in $M_p^2/(\xi_1\bpo^2)$ are negligible. Using \eq{can1}, a general term in this expansion is given  by
\be{dVsinglet}
|\Delta V_n| = \frac{(n+1)\lambda \varphi_1^n}{2\cdot 6^{n/2}\;\xi_1^3  \bpo^2 M_p^{n-6}}  ~.\ee
Substituting $\bpo^2 = N M_p^2/\xi_1$ into \eq{dVsinglet} gives
\be{dVsingletA}
|\Delta V_n| = \frac{(n+1)\lambda \varphi_1^n}{2\cdot 6^{n/2}\,N\,\xi_1^2 M_p^{n-4}}.
\ee
Thus for $n \geq 5$ the scale of unitarity-violation in the Einstein frame is
\bea \label{uviolsin}
\tilde{\Lambda}_n &\approx& \left(\frac{2\cdot 6^{n/2} \, N\,\xi_1^2}{\lambda (n+1)}\right)^{\frac{1}{n-4}} M_p \nonumber \\
& \sim & \xi_1^{2/(n-4)} M_p,
\eea
corresponding to the Jordan frame scale
\be{uviolsinJ}
\Lambda_n \sim \xi_1^{n/(2(n-4))}\bpo.
\ee
Thus $ \Lambda_n \gtrsim M_{eff} \gg \bpo$ for all $n$. Therefore the non-polynomial potential in the single real scalar case is not a source of dangerous unitarity-violation during inflation, because unitarity-violation only occurs at greater than the Planck energy in the Jordan frame.

This analysis is not possible when expanding around today's vacuum. However, we do not expect unitarity-violating effects because although the potential is non-polynomial, it tends to a flat potential in the limit of large field strength. This is the opposite behaviour from non-renormalizable interaction terms which lead to unitarity-violation, which diverge at large field strength.

\subsection{Potential sector: two real scalar fields}
In this case, the possibly dangerous terms in the potential are those containing only $\varphi_{2}$. During inflation, these have the canonically normalised form
\bea \label{Vphi2}
|\Delta V| & = & \frac{\lambda(m+1)}{4} \frac{\xi_2^m \varphi_{2}^{2m}}{\xi_1^2 M_p^{2m-4}}.
\eea
For $m \geq 3$, the unitarity-violating scale in the Einstein frame is
\be{uvphi2}
\tilde{\Lambda}_m \sim \left( \frac{\xi_1^2}{\xi_2^m}\right)^{\frac{1}{2m-4}} M_p,
\ee
corresponding to the Jordan frame scale
\be{uvphi2J}
\Lambda_m \sim \left(\frac{\xi_1}{\xi_2}\right)^{m/(2m-4)} \bpo .   
\ee
If $\xi_2 \ll \xi_1$ then $\Lambda_m \gtrsim M_{eff} \gg \bpo$ is possible and inflation can be safe with respect to unitarity-violation during inflation. However, if $\xi_2 \simeq \xi_1$, then $\Lambda_m \sim \bpo$. Therefore in the case of two real scalar fields with $\xi_1 \simeq \xi_2$, the potential term can lead to unitarity-violation in the inflaton background. However, the energy scale of unitarity-violation from the potential term is large (and so secondary) compared with that directly from the non-minimal coupling, \eq{uc3J}, which violates unitarity at $\sim \bpo/\sqrt{\xi_{2}}$ when $\xi_{1} \sim \xi_{2}$.

\section{Restoring unitarity?}
\label{dis}

So far we have considered the scale of tree-level unitarity-violation $\Lambda$ both in the vacuum and in the presence of a background inflaton field.  Depending on the specifics of the model, we have shown that this scale can be either comparable to or larger than $H_*$ and smaller than or larger than $\bpo$. Although this determines the energy scale of tree-level unitarity-violation and so whether inflation can, in principle,  be phenomenologically unaffected by the physics of unitarity-conservation, it does not address the solution of the unitarity problem. In this section, we discuss three proposed solutions to this problem. The first is a background-dependent scale of new physics, the second is the Giudice-Lee model and the third is strong coupling.

\subsection{Background-dependent scale of new physics}

If tree-level unitarity-violation is an indicator of true unitarity-violation, rather than strongly-coupled scattering, then new physics is necessary to unitarize the theory. Under what conditions can this be achieved?

An important requirement is that the scale of new physics $M_{NP}$ depends on the background field value of the inflaton. This is because the unitarity-violation scale $\Lambda_{0}$ in the present vacuum ($\bpo = 0$) is much smaller than that during inflation, and $M_{NP} \lesssim \Lambda$ must be satisfied in all vacua. This is a non-trivial requirement on a new physics sector. A notable exception is $S$-inflation with a real singlet where $\xi_2$ is chosen such that $\Lambda_0 \gg \bpo$.  In that case, there is no need for a background-dependent scale of new physics.

    The simplest interpretation of $M_{NP}$ is that it represents the mass of new particles in a sector of the theory which cancels the perturbative unitarity-violation from scattering processes in the non-minimally coupled SM. However, no explicit example of such a unitarity-completion of Higgs Inflation exists at present. The one example that has been proposed is the model of \cite{Giudice:2010ka}. We will argue below that this model is not a true completion of Higgs Inflation, but is simply the addition of an induced gravity inflation model to the SM, in which the inflaton is a heavy scalar whose coupling to the Higgs sector plays no essential role. It is important to make this clarification, as the model would otherwise represent an existence proof of models which unitarize Higgs Inflation by simply adding new particles.

If we assume that such a completion can be constructed (which is far from clear since it must cancel unitarity-violation due to a non-minimal coupling to gravity),  then the tree-level potential in the Jordan
frame {\em could} be proportional to $\bpo^{4}$, even if $M_{NP}$ is of the same
magnitude as $\bpo$ \cite{Bezrukov:2011sz}. The argument for this is that if the new
physics scale $M_{NP}$ is exactly proportional to $\bpo$, then at large $\bpo$
the only mass scale in the theory is $\bpo$ \cite{Bezrukov:2011sz}. Therefore, the
tree-level potential in the Jordan frame must be proportional to $\bpo^{4}$.
As a result, the classical Einstein frame potential will have exactly the same form as the original Higgs Inflation model. However, the connection between the low energy SM and inflation is likely to be lost unless both $\Lambda$ and $M_{NP}$ are large compared with $\bpo$. If $M_{NP}(\bpo) \approx \bpo$, then the particles in the new physics sector would contribute to the one-loop effective potential. In this case the ability to predict the spectral index based on low-energy physics will be lost and it would not be possible to know if the model is compatible with observational constraints on the spectral index. In addition, the new physics sector must have a strong effect on the Higgs sector of the SM in order that the Higgs scattering cross-section is strongly modified as $E$ approaches $M_{NP}$. This would be expected to modify the inflaton dynamics and therefore to break the connection between inflation and low energy physics, unless $\bpo$ is much smaller than $M_{NP}$.

Therefore the predictions of generalized Higgs Inflation models, their consistency with observational constraints, and even the occurrence of inflation itself, cannot be considered safe with respect to new physics corrections unless $\bpo \ll M_{NP} \lesssim \Lambda$.  This is not possible in the original Higgs Inflation model but it is possible in S-inflation with a real gauge singlet scalar when $\xi_2 \ll \xi_1$ (equivalent to $\xi_h \ll \xi_s$). However, even in $S$-inflation, it should be noted that it is a non-trivial requirement for the new physics scale to satisfy $\bpo \ll M_{NP} \lesssim \Lambda$, because the most natural scale for a background-dependent $M_{NP}$ is $\bpo$. Therefore a new physics sector appears unlikely to provide a solution to the problem of unitarity-violation in Higgs Inflation models.

\subsection{The Giudice-Lee Model}

So far the possibility of a UV complete version of Higgs Inflation with a Higgs-dependent scale of new physics has simply been assumed to be possible. One candidate is the model of Giudice and Lee (``Unitarizing Higgs Inflation", Ref. \cite{Giudice:2010ka}). This model adds a real scalar field to the SM. The Jordan frame Lagrangian is \cite{Giudice:2010ka}
\be{gl1}
 {\cal L}_{GL} = -\frac{1}{2} \left( \overline{M}^2 + \xi_{\sigma} \overline{\sigma}^2 + 2 \xi_{h}H^{\dagger}H \right)R + \frac{1}{2}(\partial_{\mu} \overline{\sigma})^2
 + |D_{\mu} H|^2 - \frac{1}{4} \kappa(\overline{\sigma}^2 -\overline{\Lambda}^2 - 2 \alpha H^{\dagger}H)^2
 - \lambda\left(H^{\dagger}H - \frac{v^2}{2}\right)^2    ~,
 \ee
where $\overline{M}$ and $\overline{\Lambda}$ are dimensionful parameters, and $\kappa,~\lambda,~ \alpha$, $\xi_{\sigma}$ and $\xi_{h}$ are dimensionless couplings. For inflation, it is necessary to choose  $\xi_{\sigma}\sim 10^4$ and $\xi_{h} \approx 1$ \cite{Giudice:2010ka}.

  The model of \eq{gl1} has the form of a two-field generalized Higgs Inflation model, but with $M_{p} \rightarrow \overline{M}$.
(We may consider the Higgs field in the unitary gauge, $H \rightarrow h/\sqrt{2}$, as the relevant unitarity-violation is due to the
$\overline{\sigma}$ and $h$ fields.) We can then use our general results to easily obtain the scale of unitarity-violation in this model. The present vacuum in this model is given by $\langle \overline{\sigma} \rangle= \overline{\Lambda}$ and $\langle h \rangle= v$. (As we consider scattering processes at $E \gg v$, we can set $v=0$.) Therefore tree-level
unitarity-violation will be given by \eq{uc3J} for the case where $\bpo =\overline{\Lambda} \approx M_{p}/\sqrt{\xi_{\sigma}}$,
$\xi_{1} = \xi_{\sigma}$ and $\xi_{2} = \xi_{h}$. This implies that unitarity is violated once $E \gtrsim M_{p}/\sqrt{\xi_{h}}$. If we then assume that $\xi_{h} \approx 1$, we find that unitarity is conserved up to the Planck scale, in agreement with the results of  \cite{Giudice:2010ka}.

   We can now understand the essential requirements for this model to work. First, the Planck mass must be dominated by the expectation value of a non-minimally coupled field; the relation $\overline{\Lambda} = M_{p}/\sqrt{\xi_{\sigma}}$ then follows. The Higgs doublet itself must have a very small non-minimal coupling, $\xi_{h} \lesssim 1$. This both prevents unitarity-violation at sub-Planck scales due to the Goldstone bosons in the Higgs doublet and also ensures that the unitarity-violation due to the $h$ and $\overline{\sigma}$ fields is also at the Planck scale or larger. It is important to emphasize that the additional singlet does not actually unitarize Higgs scattering in this theory; it is unitarized up to the Planck scale because $\xi_{h} \approx 1$.

However, is this model really a true completion of Higgs Inflation? Although it is an interesting model of non-minimally coupled inflation, we believe it is not.  Firstly, inflation is not due to the Higgs boson in this model, but due to the additional gauge singlet scalar. In addition, unlike $S$-inflation, this singlet scalar is not related to weak-scale particle physics. We can see this by restating the model in terms of a scalar field $s$ with $\langle s \rangle = 0$ in the present vacuum, $s \equiv \overline{\sigma} - \overline{\Lambda}$. \eq{gl1} becomes
\be{gl2}
 {\cal L}_{GL} =  -\frac{1}{2} \left( M_p^2 + \xi_{\sigma} (s^2   + 2 \overline{\Lambda} s )
 + \xi_{h} h^2 \right)R
+ \frac{1}{2}(\partial_{\mu} s)^2 + \frac{1}{2}(\partial_{\mu} h)^2
-\frac{1}{2} m_{s}^{2} s^2 - \frac{\lambda_{s}}{4} s^4 - \frac{\lambda_{h s}}{4} s^2 h^2
 - \frac{\lambda_{h}}{4} h^4 + \frac{\lhs}{2} \overline{\Lambda} s h^2 - \ls \overline{\Lambda} s^3  ~,
 \ee
where $M_p^2 = \overline{M}^2 + \xi_{\sigma} \overline{\Lambda}^2$, $\lambda_{s} \equiv \kappa$, $\lambda_{h s} \equiv -2 \alpha \kappa$, $\lambda_{h} \equiv \lambda + \alpha^2 \kappa$ and $m_{s}^2 = 2 \kappa \overline{\Lambda}^2$.  We see then that the mass of the singlet in this model is large, $m_{s} \approx \sqrt{\left(\kappa/\xi_{\sigma}\right)} M_{p} \sim 10^{16} \sqrt{\kappa} \GeV$.
Thus the inflaton sector has no relationship with the SM sector or weak-scale physics. \eq{gl2} has the form of an $S$-inflation model up to dimensionful couplings proportional to $\overline{\Lambda}$. Therefore the classical predictions of the model at $s \gg \overline{\Lambda}$ will be the same as $S$-inflation up to corrections of order $\overline{\Lambda}/s$, while the quantum corrections cannot be related to weak-scale particle physics.

Thus, by applying our general analysis, we are able to clarify the true nature of the model of \cite{Giudice:2010ka}. In fact, the structure of \eq{gl1} is simply that of an induced gravity inflation model \cite{induced} which has been added to the SM and which is essentially independent of the SM sector. (The coupling between the sectors, $\alpha$ in \eq{gl1}, plays no role in inflation and could be set to zero.) The Higgs sector itself plays no part in inflation. The model therefore cannot be considered a completion of a Higgs Inflation-type model in which the inflaton is part of a weak-scale particle theory.

\subsection{Strong coupling}

Given the likely difficulty of achieving a natural cancellation of Higgs Inflation unitarity-violation via a new physics sector, the remaining possibility  for unitarizing Higgs Inflation is strong coupling. In general, perturbation theory will break down before the energy of tree-level unitarity-violation is reached. Therefore it is possible that strong coupling in high-energy scattering processes will unitarize the full scattering cross-section. For the case of graviton exchange scattering of scalars, it has been shown that in the limit of large $N$, where $N$ is the number of particles in the theory, the full cross-section can remain unitary even if the tree-level cross-section breaks unitarity \cite{willenbrock}. (This possibility has been recently re-visited in \cite{dono}, which agrees with the results of \cite{willenbrock}.) $\Lambda$ would then be interpreted as a physical strong coupling scale and therefore the Higgs dependence of the unitarity-conserving physics is automatic, unlike the case of a new physics scale discussed above. In the strong coupling case, Higgs Inflation with $\Lambda \approx \bpo$ should be unaffected by the strong coupling because $\Lambda \approx \bpo \gg H_{*}$. However, it is possible that the calculation of quantum corrections to the potential could be affected by strong coupling when $\Lambda$ is close to $\bpo$ \cite{Bezrukov:2009db}. In that case $S$-inflation with a real scalar, which can have $\Lambda \gg \bpo$,  would be favoured, at least as far as retaining a predictive link with low-energy physics is concerned. For the case of complex $S$, if $\lambda_{s} ={\cal O}(0.1)$ then we expect $\Lambda \;(\approx M_{p}/\xi_{s}) \sim H_* \; (\approx \sqrt{\lambda_{s}} M_{p}/\xi_{s})$, which is disfavoured. However, if $\lambda_{s} \ll 1$, then it is possible to have $\Lambda \gg H_*$ during inflation.

It must be emphasised that strong coupling unitarization of high energy scattering in Higgs Inflation has yet to be demonstrated. Should it prove not to occur, and should no new physics completion or completion along the lines of \cite{Lerner:2010mq} be possible, then the Higgs Inflation scenario would be completely ruled out by unitarity-violation.

\section{Conclusions}
\label{conc}

We have considered the consistency with respect to unitarity-violation of generalized Higgs Inflation models with one or more scalars non-minimally coupled to gravity. By consistent we mean that the energy scales and field strengths during inflation are small compared with the energy scale of tree-level unitarity-violation, such that inflation can be analysed independently of the physics of unitarity-conservation. We first considered tree-level unitarity-violation due to graviton exchange in 2 $\rightarrow$ 2 scattering processes in the Jordan frame. In the absence of gauge fields, we find that models with two or more real scalar fields have a severe problem with
tree-level unitarity-violation during inflation if $\xi_1 \approx \xi_2$, as tree-level unitarity-violation occurs at an energy much less than the background inflaton field. This case is also relevant to the original Higgs inflation model in the limit where the energy is larger than the gauge boson masses, since the Goldstone boson fields all have the same non-minimal coupling as the Higgs inflaton.  However, the energy scale of tree-level unitarity-violation can be much larger than the background field if $\xi_2 \ll \xi_1$. Moreover, if $\xi_2 \lesssim 1/\xi_{1}$, then the energy of tree-level unitarity-violation is greater than the Planck scale, even in the present vacuum. This case is relevant to the $S$-inflation model with a real gauge singlet scalar and $\xi_s \gg \xi_h$. In this case, $S$-inflation can evade new physics corrections if the scale of new physics, such as new particle masses, is background-dependent and large compared with $\bpo$. In contrast, $S$-inflation with a {\em complex} singlet scalar is disfavoured by its low scale of tree-level unitarity-violation relative to $\bpo$. Therefore $S$-inflation with a real gauge singlet scalar is favoured over a complex singlet scalar or Higgs Inflation if unitarity-conservation is due to new physics.

The generalized Higgs Inflation model does not include gauge fields. Tree-level unitarity-violation due to Goldstone bosons only occurs when the Goldstone bosons can be considered physical, which is true once the scattering energy is larger than the gauge boson masses. Therefore the energy scale of tree-level unitarity-violation in the original Higgs Inflation model based on the Higgs doublet will be close to the background Higgs field during inflation. This disfavours the original Higgs Inflation model, as it breaks the connection between inflation dynamics, observables and low energy physics, making the model unpredictive and of unclear validity.

If the  tree-level unitarity-violation scale $\Lambda$ is instead interpreted as a strong coupling scale, where it is assumed that strong coupling can unitarize high-energy scattering processes, then both Higgs Inflation and $S$-inflation with a real singlet scalar may be safe. During inflation, the strong coupling scale in Higgs Inflation is close to $\bpo$ but can be much larger in real scalar $S$-inflation with a hierarchy of non-minimal couplings or in complex scalar $S$-inflation with $\lambda_{s} \ll 1$. Therefore if strong coupling has a large effect on the computation of the effective potential then $S$-inflation will again be favoured. Understanding the physics of strong coupling, in particular whether it can unitarize Higgs Inflation and, if so, what effect it has on the inflaton potential, is therefore an important direction for future research\footnote{In this regard, an interesting direction could be the classicalon approach to unitarization via non-perturbative dynamics \cite{dvali}.}.

    We have also considered the non-polynomial potential in the Einstein frame as a possible source of unitarity-violation during inflation. In the case of a single, real scalar field, dimensional analysis indicates that there is no dangerous unitarity-violation during inflation. In models with two or more scalars, the scale of tree-level unitarity-violation can be much larger than $\bpo$ if $\xi_2 \ll \xi_1$, but comparable to $\bpo$ if $\xi_2 \approx \xi_1$. This disfavours both Higgs Inflation and $S$-inflation with a complex scalar, but is compatible with $S$-inflation with a real scalar. Scattering via the non-polynomial potential in today's vacuum is an unsolved field theory problem. However, at least in the real scalar case, we do not expect unitarity-violation because the potential extrapolates between a simple $\phi^4$ potential and a flat, non-interacting potential.

    In general, either strong coupling or a background-dependent scale of new physics is necessary to have consistency with the small tree-level unitarity-violation scale in the present vacuum. An important exception is the case of two scalar fields with $\xi_1 \gg 1$ and $\xi_2 \ll 1$. In this case the scale of new physics in our vacuum can be larger than the inflaton field during inflation. Indeed, if $\xi_2 \lesssim 1/\xi_1$ then the scale of tree-level unitarity-violation is larger that $M_{P}$ and therefore introduces no unitarity-violation beyond that due to gravity itself. This is possible in $S$-inflation with a real scalar field if the Higgs doublet has an extremely weak non-minimal coupling.

An important question is whether a new physics completion with a background-dependent scale of new physics is actually possible. The model of \cite{Giudice:2010ka}  claims to be an example of such a completion. We have shown, however, that this is not a true completion of Higgs Inflation, but rather the addition of an essentially separate induced gravity inflation sector to the SM, with the Higgs having a conventionally small non-minimal coupling and playing no part in inflation.

Of the ways to have a unitarity-conserving Higgs Inflation model, strong coupling seems to be the best hope. No example of a Higgs-dependent new physics completion exists and even if it did, it would be non-trivial to have a model with $\bpo \ll M_{NP}(\bpo) \lesssim \Lambda$, which is necessary for Higgs Inflation to remain predictive. More generally, it is difficult to imagine how simply adding new particles could cancel unitarity-violation due to a non-minimal coupling to gravity. Moreover, the idea of introducing a new sector of heavy fields goes against the spirit of the original Higgs Inflation model, which is to explain inflation entirely with weak-scale physics. Strong coupling, on the other hand, is automatically background-dependent and, as we have shown, is less likely to significantly modify the Higgs potential during inflation. There is evidence that graviton exchange scattering may become unitary when computed to all orders \cite{willenbrock,dono}. On the other hand, should strong coupling fail to unitarize the model, then Higgs Inflation would most likely be ruled out.

In conclusion, generalized non-minimally coupled Higgs Inflation models with multiple scalar fields can be compatible with tree-level unitarity during inflation, depending on the number of scalars and their non-minimal couplings. We find that inflation models based on gauge singlet scalars generally have the best behaviour with respect to tree-level unitarity-violation. In particular, $S$-inflation with a hierarchy of non-minimal couplings can be safe and predictive during inflation. Moreover, $S$-inflation with a real scalar can remain both unitary and predictive even in the present vacuum, with no need for either strong coupling or new physics, provided there is a very strong hierarchy of non-minimal couplings.

\section*{Acknowledgements}
The work of JM is supported by the Lancaster-Manchester-Sheffield Consortium for Fundamental Physics under STFC grant ST/J000418/1. The work of RL is supported by the Academy of Finland, grant 131454.

\end{document}